\documentclass[pra, twocolumn, floatfix]{revtex4} 
\usepackage{graphicx}
\usepackage{amsmath, amsfonts, amssymb, bm}
\usepackage{braket}
\usepackage{color}
\usepackage[utf8]{inputenc}

\begin{document}

\title{ 
Radiation-field-driven ionization 
in laser-assisted slow atomic collisions  \\        
}
\author{ A. Jacob, C. M\"uller and A. B. Voitkiv}
\affiliation{ Institut f\"ur Theoretische Physik I, 
Heinrich-Heine-Universit\"at D\"usseldorf, 
\\ Universit\"atsstr. 1, 40225 D\"usseldorf, Germany }
\date{\today}

\begin{abstract} 
 
It is generally assumed that for ionization processes, which  occur in slow atomic collisions, the coupling of the colliding system to the quantum radiation field is irrelevant. Here we show, however, that -- contrary to expectations -- such a coupling can strongly influence ionization of a beam of atomic species A slowly traversing a gas of atomic species B excited by a weak laser field. 
Moreover, this coupling becomes even more important when the collision velocity increases, getting comparable or exceeding the typical electron orbiting velocities in A and B. 
Our results imply furthermore that the Breit interaction can, in fact, dominate over the Coulomb interaction at very low energies. 

\end{abstract}

\maketitle

According to Quantum Electrodynamics, atomic 
particles interact with each other by exchanging virtual photons whose four-momentum $q^\mu$ satisfies the off-shell condition $q^\mu q_\mu \neq 0$. 
The concept of virtual photons becomes especially physically appealing  
when the interaction is relatively weak and involves particles well separated in space, offering an insight into  
the basic physics of a large variety of processes ranging from 
F\"orster resonance energy transfer \cite{chromophore} and  
deexcitation processes in metallic compounds 
\cite{metal-comp}, metal oxides \cite{marpi}, rare gas dimers \cite{Ne2-dimer} and clusters \cite{Ne-cluster} to ionization reactions occurring in fast atomic collisions   
\cite{mos-1996,mos-1997,stolt}. 

In some processes their kinematics allows the interaction between particles to proceed also via the exchange of mass-shell photons with $q^\mu q_\mu = 0$. In such a case the coupling 
of the particles to the quantum radiation field   
becomes efficient and the effective range of the interaction increases dramatically that may profoundly affect characteristics of the process. 
For instance, electron-positron pair production in the collision between an extreme relativistic electron and an intense laser field proceeds mainly via the emission of an on-shell photon which is converted in the laser field into an electron-positron pair \cite{slac}. Also, the exchange of on-shell photons  
may strongly facilitate  
projectile-electron loss in high-energy collisions with atoms \cite{we-2005} and excitation of ions by high-energy electrons in the presence of an intense laser field \cite{we-2009}. 

The connecting 'bridge' between the interactions transmitted by off- and on-shell photons also exists. A prominent example is represented by   
the electromagnetic fields generated by extreme relativistic charged particles moving with velocities very closely approaching the speed of light. Such fields may become almost identical to the field of an electromagnetic wave that forms the basis of the Weizsäcker-Williams approximation of 'equivalent photons' 
\cite{weiz-willi} widely used in high-energy physics 
(see e.g. \cite{jack,bert-baur}).  

However, the above examples of processes where the interaction between atomic particles  
can be transmitted by on-shell photons 
as well as the 'bridging' regime belong to the relativistic (high-energy) domain of AMO physics. In its low-energy domain 
the situation seems to be quite different, both for weakly bound systems and collisions.  
 
For instance, important relaxation processes occurring in weakly bound systems, 
such as interatomic Auger \cite{iaad} and coulombic \cite{icd} decay, extensively studied during the last two decades in a wide variety of systems 
\cite{recent-icd-reviews}, are driven by the exchange of off-shell photons whereas the coupling to the 
radiation field plays here essentially no role 
(see e.g. \cite{nat-comm}). 
Moreover, according to textbooks 
\cite{mcdowell,sdr,mcguire} such a coupling 
is fully irrelevant for ionization and excitation processes occurring not only in slow but also quite energetic non-relativistic atomic collisions.   
 
It is, therefore, the main goal of the present communication to show that the coupling to the quantum radiation field is able to strongly influence atomic processes, even if they occur at very low energies. Moreover, it will be seen that this also implies the dominance of the Breit interaction at such low energies. (This interaction is normally regarded as important for high-energy electrons only and its  role in processes involving heavy ions has attracted much attention, see e.g. \cite{br1}.)  
As an example of such a situation, we shall consider ionization of atoms A in slow collisions with atoms B in the presence of a weak laser field resonant to electron transitions in B.  

If A and B move with respect to each other with a velocity much smaller than the typical orbiting velocities of the electrons in A and B, the impact ionization of A (or B) is strongly suppressed.  
Suppose that the ionization potential
of A is smaller than an excitation energy of a dipole-allowed
transition in B and that the A-B system is 
exposed to a weak laser field resonant to the dipole-allowed transition in B. Then 
ionization of A can proceed not only due to its direct interaction with the laser field but also 
via photoexcitation of B with its subsequent relaxation 
in which the de-excitation energy is transmitted 
to A that ionizes it.

In the case, when A and B form a weakly bound system, such a two-step ionization process was predicted in 
\cite{we-2cpi-2010} and called there (resonant) two-center photoionization (2CPI). 
It was shown in \cite{we-2cpi-2010}  
that 2CPI  
can strongly outperform the direct photoionization of A. 
A very high efficiency of 2CPI was confirmed in 
experiments on photoionization of 
the Ne-He dimer \cite{fr-exper} and Ar-Ne clusters \cite{Ar-Ne-clusters}. For 2CPI in bound systems the coupling to the radiation field turned out to be unimportant \cite{we-2cpi-2010-paper}. We shall see, however, that it is not the case for 2CPI in slow collisions.  

Atomic units ($\hbar = |e| = m_e = 1$) are used throughout unless otherwise stated. 

\vspace{0.1cm} 

Let a beam of atomic species A move with a low velocity $v$ ($v \ll 1$ a.u. $ \approx 2.18 \times 10^{8}$ cm/s) in a dilute (and cold) gas of atoms B 
in the presence of a weak monochromatic laser field with a frequency 
$\omega$  which is resonant to an electric dipole transition between the ground, 
$ u_0 $, and an excited, $ u_1 $, states of B having energies $ \epsilon_0 $ and $ \epsilon_1 $, respectively. We suppose that  
the ionization potential of A is smaller than the excitation energy $ \epsilon_1 - \epsilon_0 $  
of B. Then 
A can be ionized not only by its direct interaction with the laser field but also via the process of 2CPI  
(see Fig. 1 for illustration). 
 
\begin{figure}[h!]
\centering
\includegraphics[width=8.cm]{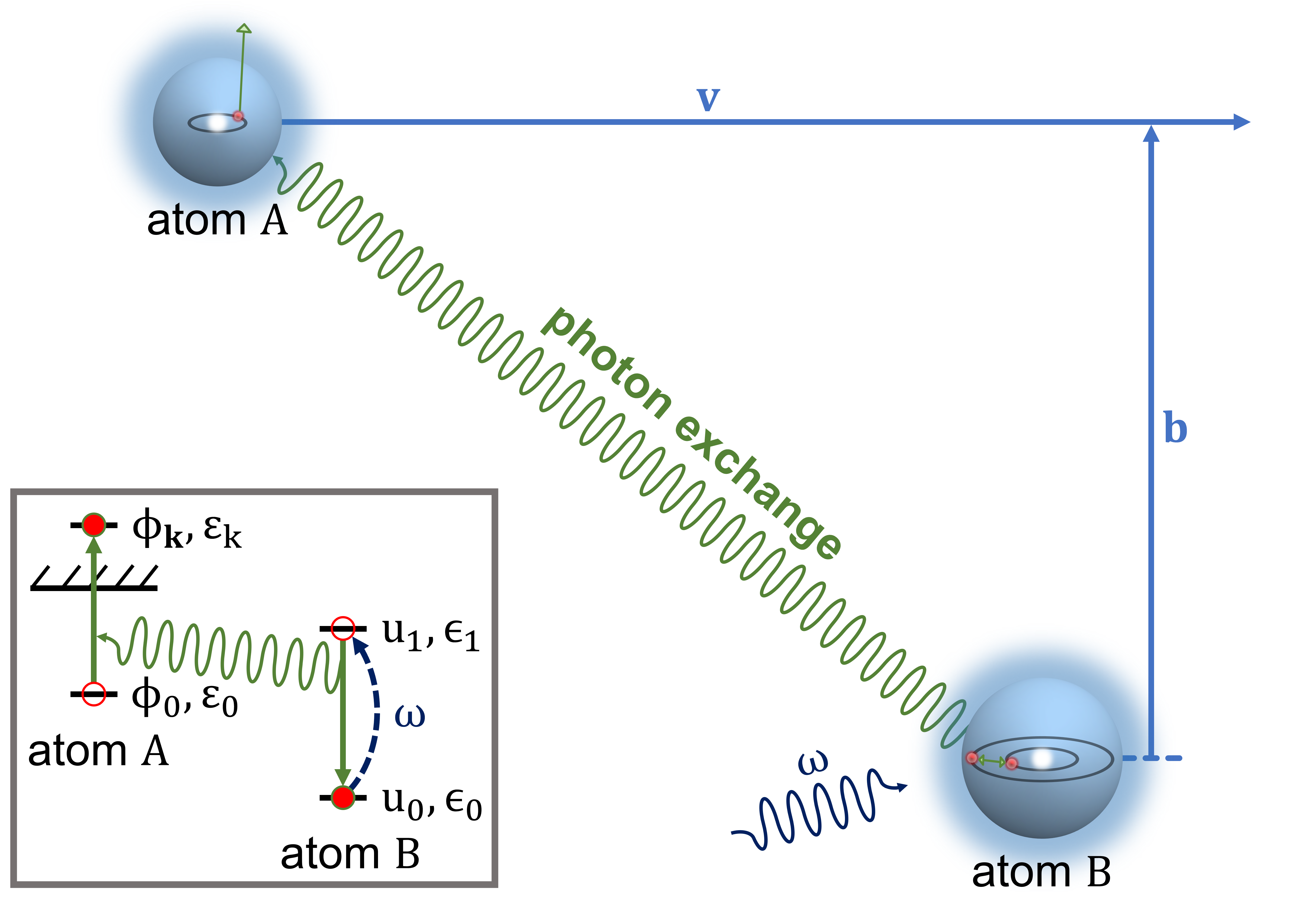}
\caption{ Scheme of collisional 2CPI. 
}
\end{figure}

The relative motion of A and B   
can be regarded as classical up to very low collision energies
($ \gtrsim 1 $ eV/u, see \cite{mcdowell}). Therefore, we shall use the semiclassical approximation in which 
the electrons are treated as quantum particles 
whereas the nuclei are supposed to move along classical 
trajectories. In distant collisions, which 
are of primary interest for the present study, 
the interaction between A and B is weak and 
their nuclei move practically undeflected.  

We choose       
a reference frame, in which B is at rest and its nucleus is taken as the origin, and A moves along a classical straight-line 
trajectory ${\bf R}(t) = {\bf b} + {\bf v} t$, where 
${\bf b} = (b_x, b_y, 0)$ is the impact parameter,  
${\bf v} = (0, 0, v)$ the collision velocity and $t$ is the time. 

Taking into account the coupling of the A-B system to the quantum radiation field and assuming
that the  
laser field (treated classically) is turned on adiabatically at $ t \to - \infty $, when A and B were in the ground states, 
the following expression can be derived for the amplitude $A_\text{2CPI}({\bf b})$ of two-center 
photoionization:   
\begin{eqnarray}
A_{\text{2CPI}}({\bf b}) \!\! & = & \! \!\frac{ -i W^B_{10}/\pi v }{ \Delta  + i \Gamma_r^{B}/2  } % \frac{-i }{\pi v } %
\! \!  \int \!\!\! d^2  {\bf q}_{\perp} %
\! \frac{ \mathcal{ P }(\bm q) \,\,  e^{- i {\bf q}_{\perp} \cdot {\bf b}} }
{ q^2 \! - \! (\varepsilon_k \! - \!  \varepsilon_0)^2/c^2 \! - \! i 0 }.      
\label{v1} 
\end{eqnarray} 
Here, 
\begin{eqnarray}
\!\!\!\mathcal{ P }({\bf q}) \!\!\! & = & \!\!\! 
\langle \phi_{\bm k} \vert e^{i {\bm q} \cdot {\bm r}} \vert \phi_0 \rangle  
\langle u_0 \vert e^{- i {\bm q} \cdot {\bm \xi}}  \vert u_1 \rangle - \frac{ 1 }{ c^2 } \times     
\nonumber \\ 
& & \!\!\!\!\!   
\langle \phi_{\bm k} \vert 
e^{i {\bm q} \cdot {\bm r}} 
(\hat{\bm p}_{\bm r} \!\! + \!\! {\bm q}/2) \vert \phi_0 \rangle \! \cdot \! \langle u_0 \vert 
e^{- i {\bm q} \cdot {\bm \xi}} 
(\hat{\bm p}_{\bm \xi} \!\! - \!\! {\bm q}/2 ) \vert u_1 \rangle,      
\label{v2} 
\end{eqnarray}
where $ \phi_0 $ and  $ \phi_{\bm k} $ are the ground state (with an energy $ \varepsilon_0$) and continuum state (with energy $\varepsilon_k$ and asymptotic momentum ${\bm k}$) of A, respectively, given in the rest frame of A, 
${\bm r}$ (${\bm \xi}$) are the coordinates  
of the 'active' electron in A (B) with respect to the nucleus of A (B), and  
$\hat{\bm p}_{\bm r} = -i {\bm \nabla_{\bm r}}$, 
$\hat{\bm p}_{\bm \xi} = -i {\bm \nabla_{\bm \xi}} $. Further, $ W^B_{10} = 0.5 \, F_0 \, \langle u_1 | {\bm e} \cdot {\bm \xi}  | 
u_0 \rangle $, where $F_0$ is the amplitude of the laser field  strength and ${\bm e}$ the polarization of the field,  $ \Delta = \epsilon_0 + \omega - \epsilon_1$ is the resonance detuning in B and 
$ \Gamma_{r}^{B }$ is the natural width of the excited state of B \cite{f1a}.  

The integration in (\ref{v1}) runs over the two-dimensional vector ${\bm q}_\perp$, which is 
the transverse part (${\bm q}_\perp \cdot {\bm v} = 0$) of the momentum transfer 
$ {\bm q}  = ({\bm q}_\perp, q_m)$ in the collision, where $q_m = (\varepsilon_k - \varepsilon_0 - \omega)/v $ is its minimum value. 
    
In the derivation of Eqs. (\ref{v1})-(\ref{v2}) 
we took into account that the characteristic velocities $v_e \sim 1$ a.u. of the 'active' electrons of A and B are much less than the speed of light $c$ ($ c \approx 137 $ a.u.) and that the collision velocity $v$ is even much smaller 
than $v_e $. However, unlike the 'standard' theories (e.g. \cite{mcdowell,sdr,mcguire}) of nonrelativistic collisions of light atomic particles (where all constituents of A and B move with velocities much less than $c$),   
the electromagnetic field transmitting the interaction between A and B is not approximated here by an instantaneous Coulomb form but is described relativistically (including the presence of the retardation term,   
$- (\varepsilon_k - \varepsilon_0)^2/c^2$, in  
Eq.(\ref{v1})). 

By decomposing the coordinate vectors ${\bm r}$  and ${\bm \xi}$ into their transverse ($\perp {\bm v} $) and longitudinal ($\parallel {\bm v} $) parts, $ {\bm r } = ({\bm r}_\perp, r_z )$ and $ {\bm \xi } = ({\bm \xi }_\perp, \xi_z )$, and denoting 
$ {\bm b} + {\bm \xi }_\perp - {\bm r}_\perp = {\bm \chi} $ we integrate in (\ref{v1}) over $d^2 {\bm q}_\perp$ obtaining  
\begin{eqnarray} 
&& \int d^2 {\bm q}_\perp 
\frac{ \exp( - i {\bm q}_\perp \cdot {\bm \chi}  ) }
{q_\perp^2 + q_m^2 - (\varepsilon_k - \varepsilon_0)^2/c^2  - i0  } 
\nonumber \\ 
 & = & \left \{ 
\begin{array}{rl}  
& 2 \pi K_0( q_r \chi); \, \, 
\vert \varepsilon_k - \varepsilon_0 - \omega \vert > \omega \, v/c \\ 
\\ 
& i \pi^2 H_0^{(1)}(q_r \chi) ; \, \, 
\vert \varepsilon_k - \varepsilon_0 - \omega \vert \leq \omega \, v/c,  
\end{array}  
\right. 
\label{v3} 
\end{eqnarray}
where $ q_r = \sqrt{ \vert q_m^2 - (\varepsilon_k - \varepsilon_0)^2/c^2 \vert} $,  
$ K_0$ is the modified Bessel function and 
$H_0^{(1)}$ is the Hankel function \cite{IStegun}.    

The functions $K_0(x)$ and $H_0^{(1)}(x)$ possess quite a different asymptotic behavior, $K_0(x) \sim \sqrt{2/\pi x} \, \, \exp(-x) $ 
and $H_0^{(1)}(x) \sim \sqrt{2/\pi x} \, \, \exp(ix - i \pi/4)$. 
Therefore, 
within the interval of electron emission energies 
$ \varepsilon_0 + \omega(1 - v/c) \leq   \varepsilon_k \leq \varepsilon_0 + \omega(1 + v/c) $  
the transition probability 
$ \vert A_{\text{2CPI}} \vert^2 $ at large impact parameters behaves 
as $1/b$, whereas outside this interval 
$ \vert A_{\text{2CPI}} \vert^2 $  
decreases at large $b$ exponentially. 

Such a large difference in the form of the transition probability means that  
within and outside the above energy interval the process of 2CPI can be regarded as proceeding via different physical mechanisms. Indeed, at  
$ \vert \varepsilon_k - \varepsilon_0 - \omega \vert \leq \omega v/c $ \, 2CPI involves the exchange of an  on-shell photon that is reflected by the real pole in the integrand of (\ref{v3}). In contrast, at 
$ \vert \varepsilon_k - \varepsilon_0 - \omega \vert >  \omega v/c $ the pole  
is complex meaning that in this case the interaction between A and B is transmitted by an off-shell photon.  
 
{\bf (i) 2CPI via coupling to the radiation field. }   
As it follows from the above consideration, the population of the interval of electron emission energies  
$\vert \varepsilon_k - \varepsilon_0 - \omega \vert \leq \omega \, v/c$ proceeds via the exchange of a photon, whose  four-momentum $ q^\mu $ satisfies the on-shell condition  
$ q_\mu q^\mu = (\omega/c)^2 - {\bm q}^2 \equiv (\omega'/c)^2 - {\bm q}'^2 = 0 $. In the rest frame of B its frequency is $\omega$ but in the rest frame of A its frequency $\omega' $ is Doppler shifted occupying the range $ \vert \omega' - \omega \vert \leq \omega \, v/c$. 

At $v \ll 1$  
this range is very narrow that enables one 
to derive a simple formula for 
its contribution $ \sigma_{\text{2CPI}}^r $   
to the total cross section for  ionization of A. Assuming for the moment that the laser field is linearly polarized along the collision velocity, 
we obtain     
\begin{eqnarray} 
\sigma_{\text{2CPI}}^r & = & \frac{ 3 \pi }{ 8} \,\,  
\frac{ \vert W^B_{10} \vert^2 }{ \Delta^2  +  
(\Gamma_r^{B})^2/4  }
\, \, 
\frac{ \Gamma_r^B \, \sigma_{ph}^A(\omega) \, b_{max}}{ v },   
\label{v4} 
\end{eqnarray} 
where $b_{max}$ ($b_{max} \gg c/\omega$) is the maximum value of the impact parameter and 
$\sigma_{ph}^A(\omega)$ is the cross section for ionization of atom A by absorption of a photon with energy $ \omega$ \cite{f3}. 
The term $ \Gamma_r^B \, \sigma_{ph}^A(\omega) 
\, b_{max} / v $ arises because  
in the case under consideration 2CPI proceeds via the coupling to the radiation field (which is also responsible for spontaneous radiative decay of the excited state of B) and the contribution of this mechanism to the rate of 2CPI at fixed positions of A  and B (and very large distances $R$ between them, 
$R \gtrsim c/\omega $) is proportional to $\Gamma_r^B \sigma_{ph}^A(\omega)/R^2$. 

Since one can show that 
$\frac{ \vert W^B_{10} \vert^2 }{ \Delta^2  +  
(\Gamma_r^{B})^2/4  } = \frac{ 1 }{ 8 \pi } \, \, 
\frac{ c }{ \omega \, \Gamma_r^{B} } \, \sigma_{sc}^B \, F_0^2$, where $\sigma_{sc}^B$ is the cross section for resonant photon scattering on B and $F_0$ is 
the amplitude of the laser field strength,  
Eq. (\ref{v4}) can be rewritten as 
\begin{eqnarray} 
\sigma_{\text{2CPI}}^r = \frac{ 3 \pi }{ 8 } \, \, 
\bigg[ n_{ph} \, c \, \frac{b_{max}}{v} \, \sigma_{sc}^B \bigg] \, \sigma_{ph}^A(\omega),   
\label{v5} 
\end{eqnarray}
where $n_{ph} = F_0^2/(8 \pi \omega)$ is the photon density in the laser field. 

Eq. (\ref{v5}) suggests that 2CPI, resulting in the electron emission into 
the energy interval $\vert \varepsilon_k - \varepsilon_0 - \omega \vert \leq \omega \, v/c$, proceeds via resonant scattering of laser photons on atoms B 
followed by absorption of the scattered photons by atoms A. The effective number of the laser photons participating in this process is   
$ \simeq n_{ph} \, c \, (b_{max}/v) \, \sigma_{sc}^B $. 

For the corresponding reaction rate,  
$ \mathcal{R}_{\text{2CPI}}^r = n_B \, v \, 
\sigma_{\text{2CPI}}^r $, we obtain 
\begin{eqnarray} 
\mathcal{R}_{\text{2CPI}}^r 
& = & \frac{ 3 \pi }{ 8 }  
\bigg[ n_B \, b_{max} \, \sigma_{sc}^B \bigg] \, 
\mathcal{R}_{\text{DPI}}^A,  
\label{v6} 
\end{eqnarray}
where $ \mathcal{R}_{\text{DPI}}^A = n_{ph} \, c \, \sigma_{ph}^A(\omega)$ is the rate for the direct photoionization of atoms A by the laser field.    

{\bf (ii) 2CPI via two-center autoionization. }   
The population of the continuum of atom A 
with energies $ \vert \varepsilon_k - \varepsilon_0 - \omega \vert > \omega \, v/c $ in distant collisions with atom B, where the electronic shells of A and B always remain well separated in space, involves the excitation of B by the laser field and a consequent 
exchange of an off-shell photon between B and A resulting in electron emission from A. 

The corresponding contribution $\sigma_{\text{2CPI}}^{nr}$ from such collisions to the cross section for ionization of A 
can be written as \cite{2cpi-ai,2cpi-in-col} 
\begin{eqnarray}
\sigma_{\text{2CPI}}^{nr} 
& = & \frac{ 9 \pi }{ 64 } \,\,  
\frac{ \vert W^B_{10} \vert^2 }{ \Delta^2  +  
( \Gamma_r^{B})^2/4 } \, \, 
\left(\frac{ c }{ \omega }\right)^4 
\, \, \frac{ \Gamma_r^B \, \sigma_{ph}^A(\omega) }{ v \, b_0^3 },     
\label{v7}
\end{eqnarray}
where $ b_0 $ is the minimum value of the impact parameter for which the electronic shells of  particles A and B do not overlap.
The last two terms in Eq. (\ref{v7}) point to the fact that in this case  
2CPI involves two-center autoionization  
whose rate $\Gamma_a $ at large distances $R$ between A and B 
($R \gg 1$ a.u.) is proportional to  
$ (c/\omega)^4 \, \Gamma_r^B \, \sigma_{ph}^A(\omega) /R^6  $ (see e.g. \cite{iaad}, \cite{we-iaad-2020}).  

The cross section 
$ \sigma_{\text{2CPI}}^{nr} $ can also be presented  as   
\begin{eqnarray}
\sigma_{\text{2CPI}}^{nr} & = & 
\frac{ 9 \pi }{ 64 } \, n_{ph} \, c \, \sigma_{sc}^B \, \frac{ c }{ \omega v } \, \bigg(\frac{ c }{ \omega b_0 }\bigg)^3 \sigma^A_{ph}(\omega). 
\label{v8}
\end{eqnarray}
The corresponding reaction rate, 
$\mathcal{R}_{\text{\text{2CPI}}}^{nr} = n_B \, v \, 
\sigma_{\text{2CPI}}^{nr} $, reads  
\begin{eqnarray}
\mathcal{R}_{\text{\text{2CPI}}}^{nr} & = & \frac{ 9 \pi }{ 64 } \, n_B \,  \sigma_{sc}^B \, \frac{ c }{ \omega } \, 
\mathcal{R}_{ \text{DPI} }^A \, \bigg(\frac{ c  }{ \omega b_0 }\bigg)^3. 
\label{v9}
\end{eqnarray}  

{\bf (iii) } 
The relative effectiveness of the above two mechanisms   
can be characterized by the ratio 
\begin{eqnarray} 
\eta = \frac{ \mathcal{R}_{\text{\text{2CPI}}}^r }{ 
\mathcal{R}_{\text{\text{2CPI}}}^{nr} }   
= 
\frac{ 8 }{ 3 } \frac{ b_{max} b_0^3 }{ (c/\omega)^4 }.   
\label{v10} 
\end{eqnarray}
This ratio strongly depends on the amount of energy transferred from B to A, suggesting that 2CPI in systems, where it proceeds with a large energy transfer, is likely to be dominated by the exchange of an on-shell photon whereas the opposite will be true if the energy transfer is small \cite{f4}. 

For instance, simple estimates show that 2CPI in slow collisions, where species B are represented by Rb atoms driven by a laser field resonant 
to its 5s$_{1/2}$--5p$_{3/2}$ transition  
($\omega \approx 1.59$ eV), will be dominated by the two-center autoionization for any realistic choice of $b_{max}$. On the other hand, if B are He atoms in a laser field resonant to the $1 \, ^1S_0$--$2 \, ^1P_1$ transition ($\omega \approx 21.1$ eV), 2CPI in slow collisions will be dominated by the exchange of an on-shell photon provided $b_{max} \gtrsim 1$ mm.  

{\bf (iv) 2CPI and Breit interaction. } 
Slowly moving charged particles  
normally interact with each other mainly via the (instantaneous) Coulomb force. When the interacting particles are electrons the leading correction to the latter is given by the (generalized) Breit interaction \cite{gbi}. 

The contributions of these interactions to 2CPI can be made explicit by rewriting the amplitude (\ref{v1}) 
as a sum of two terms,  
\begin{eqnarray}
A_{\text{2CPI}}({\bf b}) & = &  
A^{\text{coul}}_{\text{2CPI}}({\bf b}) + A^{\text{breit}}_{\text{2CPI}}({\bf b}).            
\label{v-s-1-IV} 
\end{eqnarray}
Here, the term 
\begin{eqnarray}
A^{\text{coul}}_{\text{2CPI}}({\bf b}) \!\! & = & 
 \! \!\frac{ -i W^B_{10}/\pi v }{ \Delta  + i \Gamma_r^{B}/2  } 
 \int \!\! d^2  {\bf q}_{\perp} \, \, 
 e^{- i {\bf q}_{\perp} \cdot {\bf b}}  
\nonumber \\ 
&& 
 \times \frac{  \langle \phi_{\bm k} \vert e^{i {\bm q} \cdot {\bm r}} \vert \phi_0 \rangle }{ {\bm q}^2 }     
\label{v-s-2-IV} 
\end{eqnarray}
arises due to the exchange of time-like and longitudinal photons transmitting the (unretarded) Coulomb interaction  between the electrons of atoms A and B. 
The other term, given by 
\begin{eqnarray}
A^{\text{breit}}_{\text{2CPI}}({\bf b}) \!\! & = & \! \!\frac{ i W^B_{10}/\pi v }{ \Delta  + i \Gamma_r^{B}/2  } 
 \int \!\! d^2  {\bf q}_{\perp} \, \, e^{- i {\bf q}_{\perp} \cdot {\bf b}} 
\nonumber \\ 
&&   
\times \frac{ \mathcal{ Q }(\bm q)/c^2 }
{ {\bm q}^2 - \frac{ (\varepsilon_k - \varepsilon_0)^2}{c^2} - i 0 }          
\label{v-s-3-IV} 
\end{eqnarray}
with 
\begin{eqnarray}
\!\!\!\mathcal{ Q }({\bf q}) \!\!\! & = & \!\!\!  
\langle \phi_{\bm k} \vert 
e^{i {\bm q} \cdot {\bm r}} 
(\hat{\bm p}_{\bm r} \!\! + \!\! {\bm q}/2) \vert \phi_0 \rangle \! \cdot \! \langle u_0 \vert 
e^{- i {\bm q} \cdot {\bm \xi}} 
(\hat{\bm p}_{\bm \xi} \!\! - \!\! {\bm q}/2 ) \vert u_1 \rangle +  
\nonumber \\ 
& & \! \! \! \frac{ \varepsilon_k \! - \! \varepsilon_0 }{ {\bm q}^2 } 
\langle \phi_{\bm k} \vert 
e^{i {\bm q} \cdot {\bm r}} \! 
({\bm q} \! \cdot \! \hat{\bm p}_{\bm r} \!\! + \!\! {\bm q}^2/2) \vert \phi_0 \rangle \! \langle u_0 \vert 
e^{- i {\bm q} \cdot {\bm \xi}} \vert u_1 \rangle,           
\label{v-s-4-IV}
\end{eqnarray} 
appears due to the exchange of transverse photons responsible for the Breit interaction between the electrons of A and B. 

The integrand in the expression 
for $A^{\text{coul}}_{\text{2CPI}}$ 
(see Eq. (\ref{v-s-2-IV})) does not contain a pole on the real axis of $q_\perp$ whereas  
the integrand in the expression for $A^{\text{breit}}_{\text{2CPI}}$ (Eq. (\ref{v-s-3-IV})) does have a real pole if $q_m^2 \leq (\varepsilon_k-\varepsilon_0)^2$. 
When the latter condition (which is equivalent to 
$ \vert \varepsilon_k - \varepsilon_0 - \omega \vert \leq v \omega/c $) is fulfilled, the A-B system couples 'resonantly' to the radiation field. As a result, the Breit interaction is transmitted by on-shell photons that strongly enhances its efficiency and can  make this interaction dominant. This point is quite  spectacular since the role of the Breit interaction  usually remains relatively modest in atomic collision physics, even for processes involving high-energy electrons. 

{\bf (v) 2CPI versus direct photoionization. }  
The process of 2CPI competes with the direct photoionization (DPI) of atoms A by the laser field. The relative effectiveness of these two processes can be described by the ratio of the total rate 
$ \mathcal{R}_{\text{\text{2CPI}}} = \mathcal{R}_{\text{\text{2CPI}}}^r + 
\mathcal{R}_{\text{\text{2CPI}}}^{nr}$ for 2CPI and 
the rate $\mathcal{R}^A_{ \text{DPI} }$ for direct photoionization:  
\begin{eqnarray} 
\zeta & = & \frac{ \mathcal{R}_{\text{\text{2CPI}}} } { 
\mathcal{R}^A_{ \text{DPI} } }  
= \frac{ 3 \pi }{ 8 } \frac{ b_{max} }{ \Lambda_{rad}^B } + 
\frac{ 9 \pi }{ 64 } \, 
\frac{ (c/\omega)^4 }{ \Lambda_{rad}^B \, b_0^3 },    
\label{v11} 
\end{eqnarray} 
where $ \Lambda_{rad}^B = 1/(n_B \,  \sigma_{sc}^B) $ is the mean free path of the radiation in the gas of atoms B. 

It follows from (\ref{v11}) 
that 2CPI driven by the exchange of an off-shell photon can be more efficient than DPI provided the photon energy is sufficiently small (such that the factor $(c/\omega b_0)^3$ becomes very large). 

In the opposite case, when the energy transfer between B and A is relatively large and 2CPI is already dominated by the exchange of an on-shell photon, the  
value of $\zeta$ will be determined by the first term of the sum in Eq. (\ref{v11}) which involves the ratio of two distances, $b_{max}$ and $\Lambda_{rad}^B$.  The magnitude of $b_{max}$ depends on the size of the target gas of atoms B and/or the size of the beam of atoms A. However, its maximum value obviously cannot considerably exceed $\Lambda_{rad}^B$ and, in the case $b_{max} = \Lambda_{rad}^B$, we obtain 
$ \zeta = 3 \pi/8 $. This means that the resonant scattering of laser photons by atoms B with their consequent absorption by atoms A can more than double the ionization rate for atoms A. 

Besides, if the laser field is circularly polarized propagating in the direction parallel/antiparallel to 
the collision velocity, the ratio 
$ \zeta = \mathcal{R}_{\text{\text{2CPI}}}/ \mathcal{R}^A_{ \text{DPI} } $ becomes even somewhat larger, 
\begin{eqnarray} 
\zeta = \frac{ 9 \pi }{ 16 } \frac{ b_{max} }{ \Lambda_{rad}^B } + 
\frac{ 27 \pi }{ 128 } \, 
\frac{ (c/\omega)^4 }{ \Lambda_{rad}^B \, b_0^3 },    
\label{v12} 
\end{eqnarray}  
and in the case, when this ratio is dominated by the exchange of an on-shell photon, 
the ionization rate for atoms A almost triples. 

We note that 2CPI in slow distant collisions of two species A and B was considered in \cite{2cpi-in-col} by regarding the interaction between them as instantaneous, 
i.e. being transmitted by off-shell photons only.   
The textbooks on atomic collisions  
(see e.g. \cite{mcdowell,sdr,mcguire}) strongly suggest that such an approach is just appropriate 
to describe slow collisions of light atomic species where all the particles involved (electrons and nuclei) move with velocities orders of magnitude smaller than the speed of light.   

However, as we have seen, a more complete treatment of 2CPI has to take into account the coupling of the A-B system to the radiation field which may become very efficient, enabling the interaction to be transmitted by on-shell photons and profoundly modifying the process of 2CPI.  

{\bf (vi) Ionization at larger collision velocities. }
Moreover, as a more general consideration shows, expressions 
for the cross section $\sigma_{\text{2CPI}}^r $ 
and the rate $ \mathcal{R}_{\text{2CPI}}^r $ 
for ionization of A via the exchange of on-shell photons, 
given by Eqs. (\ref{v5}) and (\ref{v6}), respectively, 
remain valid as long as the collision velocity is 
much less than the speed of light. This, in particular, means that 
the rate $ \mathcal{R}_{\text{2CPI}}^r $ does not depend on the collision velocity up to impact energies of $\sim 10$ MeV/u. Since at impact energies $\gtrsim 0.5$ MeV/u the strength of all  ionization mechanisms, driven by the exchange of off-shell photons, 
rapidly decreases \cite{sdr}, 
the coupling to the radiation field may become  
in this case even more important. 
 
{\bf (vii)} Let us now consider a possible experiment on 2CPI involving the most simple atoms, H and He. Let a beam of slow H atoms  
penetrate a gas of cold He atoms in the presence of a very weak (intensity $\lesssim 10^2 $ W/cm$^2$) monochromatic laser field of linear polarization, which propagates perpendicular to the collision velocity and is resonant to the $1 \, ^1S_0$--$2 \, ^1P_1$ transition in He ($\omega \approx 21.1$ eV). By choosing $b_{max} = 5$ mm and taking a conservative estimate for the parameter 
$b_0$ ($b_0 = 3$ a.u.) we obtain from Eq. (\ref{v10}) $\eta \approx 7 $ which means that 2CPI is strongly dominated by the exchange of an on-shell photon. 

The total cross section $\sigma_{sc}^B$  for photon scattering on He at the exact resonance is $\approx 1.63 \times 10^{-11}$ cm$^2$ and the mean free path $\Lambda_{rad}^B$ for the radiation in the He gas will be equal to $b_{max} = 5$ mm at 
the gas density of $n_B \approx 1.23 \times 10^{11}$ cm$^{-3}$. Under such conditions, as Eq. (\ref{v11}) shows, 2CPI results in about the same ionization rate for H atoms as the direct photoionization.  

In order to minimize the Doppler broadening of the spectral line caused by the thermal motion of He  atoms in the target gas, the temperature $T$ of the latter should be $T \lesssim 0.1$ K that will make the Doppler broadening significantly smaller than  the natural linewidth 
$\Gamma_r^B \approx 1.23 \times 10^{-6}$ eV of 
the $2 \, ^1P_1 \to 1 \, ^1S_0$ transition in He. 
In addition, there might also occur pressure broadening of the spectral line caused by atomic collisions or quasi-static atom-atom interactions in the target medium. However, in a dilute gas ($n_B \lesssim 10^{13}$ cm$^{-3}$) such effects are usually much smaller than the natural linewidth and 
can be neglected. 
 
A more sophisticated experiment 
can also be envisaged, in which a gas target of cold atoms B is irradiated by a weak laser field but the beam of atoms A passes close by without penetrating the target and without interacting with the laser field. In such a case only 2CPI proceeding via the coupling of the A-B system to the radiation field can contribute to ionization of A.  
 
In conclusion, exploring ionization of atoms A in slow collisions with atoms B excited by a weak laser field, we have shown that this process can be strongly enhanced due to the coupling of the A-B system to the quantum radiation field and that this coupling becomes even more important when the collision velocity increases (still remaining much less than the speed of light). The basic reason for this enhancement is that the coupling to the radiation field enables the interaction between A and B to proceed via the exchange of an on-shell photon, this way dramatically increasing its effective range. 

Considering the process of 2CPI as 
a 'competition'  
between the Coulomb and Breit interactions  
we showed that 
the latter can be dominant even at very low energies. 

The predicted effects can be tested in
an experiment in which a beam of slow 
hydrogen atoms -- or some other species, 
e.g. Mg$^{+}$ ions having a binding energy 
of $\approx 15$ eV --   
penetrates (or passes close by)  
a dilute ($n_B \simeq 10^{11}$ cm$^{-3}$) target of cold He atoms 
in the presence of a weak ($\lesssim 10^2$ W/cm$^2$) laser field resonant to the 
$1 \, ^1S_0$--$2 \, ^1P_1$ transition   
in He.


\begin{thebibliography}{33}
\bibitem{f1a} We note that the 'weak laser field' means $ | W^B_{10} | < \Gamma_r^{B}/2 $.
\bibitem{f3} One should note that the quantity   
$ \sigma_{\text{2CPI}}^r $ depends on the collision geometry ($b_{max}$) and, therefore, is not a 'normal' cross section.
\bibitem{2cpi-ai} The cross section $\sigma_{\text{2CPI}}^{nr}$ was obtained in 
\cite{2cpi-in-col}. Here we presented 
this cross section in a more convenient form given by Eq. (\ref{v7}).
\bibitem{f4}  We note that the correspondence between the two mechanisms of 2CPI closely resembles that between spontaneous radiative decay and autoionization in atoms and ions where the former dominates transitions with a large energy release whereas the latter is stronger when the energy release is small.
\end{thebibliography}
\end{document}